\documentclass[preprintnumbers,aps,pra,groupedaddress]{revtex4}

\begin{document}
\draft
\title{Reply to Comment on \textquotedblleft Neutrino oscillations originate
from virtual excitation of Z bosons\textquotedblright\ and \textquotedblleft
Neutrinos produced from $\beta$ decays of neutrons cannot be in coherent
superpositions of different mass eigenstates\textquotedblright }
\author{Shi-Biao Zheng}
\thanks{E-mail: t96034@fzu.edu.cn}
\address{College of Physics and Information Engineering, Fuzhou University,\\
Fuzhou 350108, China}

\begin{abstract}
In this reply, I point out that the comment by Cline (arXiv:2410.05826) on
my manusripts (arxiv:2407.00954 and arxiv:2410.03133) has overlooked the
critical fact that the quantum entanglement between the neutrino's mass and
other degrees of freedom would destroy the quantum coherence between the
mass eigenstates.
\end{abstract}

\vskip0.5cm

\narrowtext

\maketitle

\bigskip In a recent manuscript [1], I have proved that the electron
antineutrino created by the $\beta $ decay of a neutron cannot be in a
coherent superposition of different mass eigenstates. In another manuscript
[2], I have proposed a new mechanism, where the transformations among the
neutrino's flavors are induced by the virtual excitation of the Z bosonic
field that can connect different neutrino flavors. In the comment by Cline
[3], it was stated that the claims in my manuscripts are incorrect. As
detailed below, the main statements of the comment are actually wrong.

Firstly, it should be pointed out that the author of the comment has
seriously misinterpreted the first point of my manuscript by saying ``{\sl To
prove the first point, the author assumes that the neutrino in question is
in an eigenstate of momentum. If it is a superposition of different mass
eigenstates, then clearly it cannot also be an eigenstate of energy. However
the author believes it should be an eigenstate of energy as well.}" I have
never meant that the neutrino should be in an eigenstate of energy. What I
said in Ref. [2] is ``{\bf However, I find that the energy conservation law
requires these mass eigenstates, if they exist, to be entangled with
distinct joint energy eigenstates of the other particles produced by the
same weak interaction as the neutrino}."

Then the author commented ``{\sl In reality, a neutrino emitted in a weak
interaction is a flavor state, which is a superposition of mass eigenstates.
Whether the states in this superposition have exactly the same energy,
momentum, or neither, is irrelevant to the fact that they will oscillate.}"
This statement has actually overlooked the fact that the entanglement
between the neutrino's mass and other degrees of freedom (e.g., momenta) is
sufficient to destroy the quantum coherence among the mass eigenstates.
These degrees of freedom act as a which-path detector to encode the
information about which mass eigenstate the neutrino is in. The quantum
entanglement between the interfering particle and the which-path detector
would inevitably deteriorate the quantum coherence of the interfering
particle. In my manuscript, I have unambiguously demonstrated the three
neutrino mass eigenstates, if there exist, would be left in a classical
mixture when the momentum degrees of freedom are traced out. The
entanglement-interference relation, which is a consequence of Bohr
complementarity, has been theoretically investigated [4-6], and
experimentally demonstrated in a number of physical systems [7-13]. It
should be noted that the decoherence caused by entanglement with other
degrees of freedom is unconditional; it does not matter that these degrees
of freedom are actually not measured, as highlighted in the experimental
paper of Ref. [7]. As the quantum coherence is destroyed by quantum
entanglement, no oscillations can occur. It should be further pointed out
that the concepts of quantum mechanics are valid even at the subatomic
level, exemplified by a recent experiment, where entanglement between top
quarks was observed [14]. Actually, quantum field theory is based on
relativistic quantum mechanics, and certainly could not be used to negate
the concepts or principles of quantum mechanics.

I agree with the statement ``{\sl This age-old issue has been discussed
widely in the literature; see for example [2-9]}", but which does not mean
that this issue has been adequately solved in the literature. As a matter of
fact, the issue of entanglement-induced decoherence has been largely
overlooked and has not been fully understood in these investigations. For
example, in Ref. [2] of the comment (Phys. Atom. Nucl. 72, 1363;
arXiv:0905.1903), it was said ``{\sl The subsequent disentanglement, which is
necessary for neutrino oscillations to occur, is assumed to be due to the
interaction of these accompanying particles (such as e.g. electrons or muons
produced in decays of charged pions) with medium.}" This statement correctly
recognizes that the entanglement would destroy quantum coherence of the
neutrino mass eigenstates necessary for oscillations. However, the claim
that disentanglement can make oscillations occur is not correct. Due to
quantum entanglement between the neutrino and the accompanying particles,
the three mass eigenstates are essentially in a classical mixture when the
degrees of freedom of the accompanying particles are traced out. There is no
way to turn such a classical mixture into a quantum superposition during the
neutrino's free propagation, as which would violate the principle of entropy
increase. When the accompanying particles are disentangled from the neutrino
by interaction with the medium, the entanglement of the accompanying
particles with the neutrino is transferred to the medium, but cannot
disappear.

Another main statement of the comment is ``{\sl Unfortunately he assumed an
incorrect starting point, including tree-level flavor changing
neutral-current interactions of Z with the neutrinos. This of course is not
what nature has chosen; therefore the results found in [1] are incorrect.}"
This statement seems to be quite unreasonable. What is the nature's favorite
choice? According to the previous proposal, an additional element (e.g, a
very heavy neutrino that is beyond the standard model, as proposed in the
seesaw mechanism [15]) needs to be introduced to account for the tiny
neutrino masses. This is unnecessary in the present proposal. According to
Occam's Razor: Entities should not be multiplied unnecessarily, which
proposal does nature prefer to choose? Finally, I would like to point out
that the presence of off-diagonal terms in an neutral-current interaction
was first proposed by Wolfenstein [16], where the effective Hamiltonian was
obtained for the neutrino propagating in matter. In my manuscript [2], I
showed that these terms can also appear in the vacuum, where the Z bosonic
field is also present.

\end{document}